\documentclass[submission,copyright,creativecommons]{eptcs}
\providecommand{\event}{MARS 2022} 
\usepackage{breakurl}             
\usepackage{underscore}           
\usepackage{graphicx}
\usepackage[acronym]{glossaries}

\title{Stateful to Stateless\\ Modelling Stateless Ethereum}
\author{Sandra Johnson
\institute{ConsenSys Software Inc, Australia\thanks{Corresponding author}}
\institute{ARC Centre of Excellence for\\ Mathematical \& Statistical Frontiers (ACEMS) \\
QUT, Brisbane, Australia}
\email{sandra.johnson@consensys.net}
\and
David Hyland-Wood  
\institute{ConsenSys Software Inc, Australia}
\institute{University of Queensland\\
Brisbane, Australia}
\email{d.hylandwood@uq.edu.au}
\and
Anders L Madsen  
\institute{HUGIN EXPERT A/S, Aalborg University\\
Denmark}
\email{anders@hugin.com}
\and
Kerrie Mengersen 
\institute{School of Mathematical Sciences and \\Centre for Data Science \\
Queensland University of Technology (QUT)\\
Brisbane, Australia}
\email{k.mengersen@qut.edu.au}
}
\def\titlerunning{Modelling Stateless Ethereum}
\def\authorrunning{S. Johnson, D. Hyland-Wood, A.L. Madsen \& K. Mengersen}
\def\copyrightholders{Johnson, Hyland-Wood, Madsen \& Mengersen}
\begin{document}
\maketitle

\newacronym{bn}{BN}{Bayesian network}
\newacronym{cpt}{CPT}{conditional probability table}
\newacronym{dag}{DAG}{directed acyclic graph}
\newacronym{dbn}{DBN}{Dynamic Bayesian network}
\newacronym{doobn}{DOOBN}{Dynamic object-oriented Bayesian network}
\newacronym{dos}{DoS}{Denial of Service}
\newacronym{io}{IO}{Input/Output}
\newacronym{oobn}{OOBN}{Object-oriented Bayesian network}
\newacronym{stf}{STF}{state transition function}

\begin{abstract}
The concept of `Stateless Ethereum' was conceived with the primary aim of mitigating Ethereum's unbounded state growth. The key facilitator of Stateless Ethereum is through the introduction of `witnesses' into the ecosystem. The changes and potential consequences that these additional data packets pose on the network need to be identified and analysed to ensure that the Ethereum ecosystem can continue operating securely and efficiently. In this paper we propose a Bayesian Network model, a probabilistic graphical modelling approach, to capture the key factors and their interactions in Ethereum mainnet, the public Ethereum blockchain, focussing on the changes being introduced by Stateless Ethereum to estimate the health of the resulting Ethereum ecosystem. We use a mixture of empirical data and expert knowledge, where data are unavailable, to quantify the model. Based on the data and expert knowledge available to use at the time of modelling, the Ethereum ecosystem is expected to remain healthy following the introduction of Stateless Ethereum.
\end{abstract}
 
\section{Introduction}
Ethereum~\cite{wood2016a} is the largest blockchain after Bitcoin~\cite{nakamoto2008}. When a new node wishes to join 
Ethereum mainnet, the public Ethereum blockchain, it has to acquire all of the blockchain history, starting from the first block up to the most recently added block. The storage space required depends largely on whether the node is operating as a `full' node, `archive' node, `miner', or `light client', and to some extent on the chosen Ethereum client software that is used to run the node. 

The Ethereum world state contains all Ethereum accounts, their balances, deployed program code known as ``smart contracts'', and associated storage\footnote{\href{https://ethereum.org/en/developers/docs/}{Ethereum Development Documentation}}. New accounts are continually being added and new smart contracts are being deployed. Therefore, by design, the state size of Ethereum keeps growing ad infinitum, leading to increased cost and time to sync to the network, which in turn leads to a less diverse ecosystem of Ethereum nodes, and slower transaction processing and block verification times.

Vitalik Buterin recognised these issues back in 2017 when he first introduced the concept of Stateless Ethereum suggesting a protocol transformation based on \glspl{stf}~\cite{Buterin17}. The key mechanism of Stateless Ethereum is the creation of block ``witnesses'' so that when clients receive validated blocks from miners, they will also receive their corresponding witness. Each witness contains a set of Merkle branches that are authenticated against the state root and contains all the data required to execute the transactions contained in that particular block, i.e. all the information required to prove the validity of the block and to perform the state transition. 

With witnesses comes increased traffic on the network, because more data packets will be passed around the network. We therefore need to assess the impact that this may have on the network to ensure that the Ethereum ecosystem continues to operate securely and efficiently in this new, altered ecosystem. Since Ethereum mainnet is a public blockchain, it provides a rich data source for modelling and analysis. 

\section{Modelling Approach}
\label{sec:model}

Ethereum is a complex environment, and modelling such an ecosystem is very challenging. From the myriad of processes and interactions, we need to extract those that are most likely to be affected in some way by introducing statelessness into this equilibrium. Once we have a clearer idea of the key processes that may be impacted, we can start building a model of Ethereum as it is today, and include those processes that are specific to Stateless Ethereum. 

When building a model we aim to capture what we know, and facilitate the exploration of that which we do not know. Therefore, we consult with experts to identify `key factors' that best capture the behaviour of the system, or parts thereof. Representing expert knowledge in the model is crucial, not only to identify those processes that may be affected by the changes, but to help achieve the necessary balance between system detail and model compactness, to avoid unneccesary model clutter. 

From the plethora of modelling techniques that are available, we decided on a \gls{bn}, a probabilistic graphical modelling approach, for several reasons.

First, the visual nature of a \gls{bn} facilitates ease of interaction by all stakeholders with the model. Secondly, in a BN uncertainty is captured in the model using joint probability distributions across all the variables and their interactions with other variables. Thirdly, this modelling approach is well suited to modelling complex systems like Ethereum, and lastly, diverse data sources can be used to quantify the model, including expert knowledge, empirical data, model output, published and grey literature.

A \gls{bn} is a probabilistic representation of variables and their directed relationships \cite{Pearl1988}, modelled as a \gls{dag} comprising nodes representing variables, directed links (edges) between nodes representing relationships between the variables, and conditional probability distributions for each variable representing the nature of these relationships. The two key characteristics of \glspl{bn} are the Markov property and $d$-separation \cite{Pearl1988, Lauritzenetal1990, JensonNielson2007}. The Markov property allows that there are no other direct dependencies in the network over and above those already represented by directed links in the network. D-separation, or `directional' separation is a graphical criteria for reading statements of conditional dependence and independence from the structure of the \gls{bn}.

Since \glspl{bn} were first introduced by Pearl in 1988  \cite{Pearl1988}, many papers and textbooks have been written on this modelling technique, including theoretical and applied aspects of \glspl{bn} and \gls{bn} learning \cite{Pearl1988, Cowelletal2007, JensonNielson2007, KoskiNoble2009}. \glspl{bn} are now common in a very wide range of applications such as medicine, computing, natural sciences and engineering \cite{Pourretetal2008, ScutariDenis2015}, as well as in more general fields such as decision analysis \cite{Smith2010} and risk assessment \cite{FentonNeil2019}.

Moreover, several variations have emerged to better suit requirements of systems being modelled. For example, a \gls{dbn} to better facilitate systems that change over time, and an \gls{oobn}  which is useful to model large, complex hierarchical systems \cite{KollerPfeffer1997}.  

\glspl{oobn} provide a more modular approach to modelling consisting of subnetworks and interfaces. Each subnetwork can be a complete \gls{bn} model in its own right, or a network fragment for repeating patterns in the model. The interface(s) between these subnetworks define the information flow between them. The subnetworks can also be nested to represent a hierarchical system. A key advantage of this approach is the ability to develop the objects in parallel, using the interface to protect against internal changes in these \gls{bn} sub-models \cite{Johnson2013}.  Building a \gls{bn} is an iterative process \cite{Johnson2012a, Marcot2007}.

Sensitivity analysis can be used to assess the sensitivity of specific variables to variations in the evidence entered into the network (evidence sensitivity) and to variations in the values of the parameters (parameter sensitivity) \cite{Pourretetal2008, ScutariDenis2015}. Popular measures of evidence sensitivity are entropy, which can be interpreted as the average additional information necessary to specify an alternative, and mutual information, which represents the extent to which the joint probability of two variables differs from what it would have been if they were independent \cite{Pearl1988, KorbNicholson2011}. Parameter sensitivity metrics include an evaluation of the first derivative of the sensitivity ratio of the parameter of interest and the target probability \cite{Pollinoetal2007}.

An object-oriented \gls{bn} model was chosen for the Stateless Ethereum model, to best deal with the complexity of the ecosystem and take advantage of sub-models that can be run independently.

\section{Stateless Ethereum Model Structure}
To the best of our knowledge, a probabilistic modelling approach such as \glspl{bn} has not previously been used to model Ethereum nor Stateless Ethereum.

\paragraph{Stateless Ethereum \gls{oobn}}

\begin{figure}[htb] 
\centering
\includegraphics[width=0.6\linewidth]{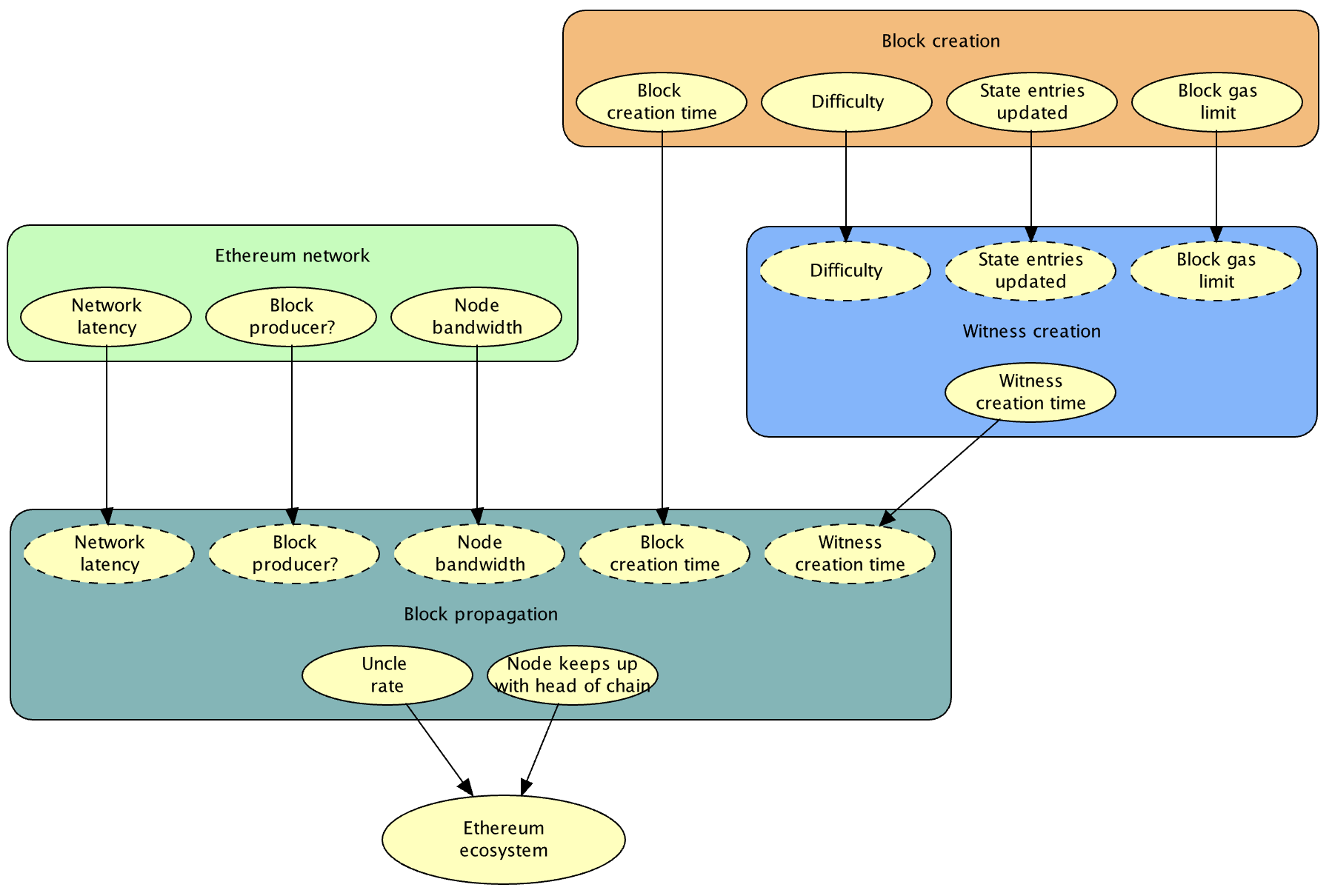}
\caption{High level model of Stateless Ethereum showing sub-models and interfaces}
\label{fig:stateless}
\end{figure}

Figure~\ref{fig:stateless} on page~\pageref{fig:stateless} shows a high-level view of the complete model, comprising of four sub-models: \textit{Block creation}, \textit{Witness creation}, \textit{Ethereum network}, and \textit{Block propagation}. Each rounded rectangle represents an \gls{oobn} sub-model. 
In each of the sub-models, we captured the key variables and their interactions in Ethereum mainnet and focussed on the changes introduced by Stateless Ethereum. The flow of information through the sub-models are via input (ellipse with broken line) and output nodes (ellipse with solid line), the former acts as a placeholder and the latter means the node is visible to other sub-models \cite{KollerPfeffer1997, Johnson2013}. In addition to input and output nodes, an \gls{oobn} can have private nodes. These key factors may be added, changed or deleted without affecting the other sub-models, other than through the predefined interface nodes (output and input nodes) \cite{Johnson2013}.

Each sub-model is explained in more detail below.

\paragraph{Ethereum network \gls{oobn}}
\label{sec:ethereumnodeoobn}

\begin{figure}
\centering
\begin{minipage}{.39\linewidth}
  \includegraphics[width=\linewidth]{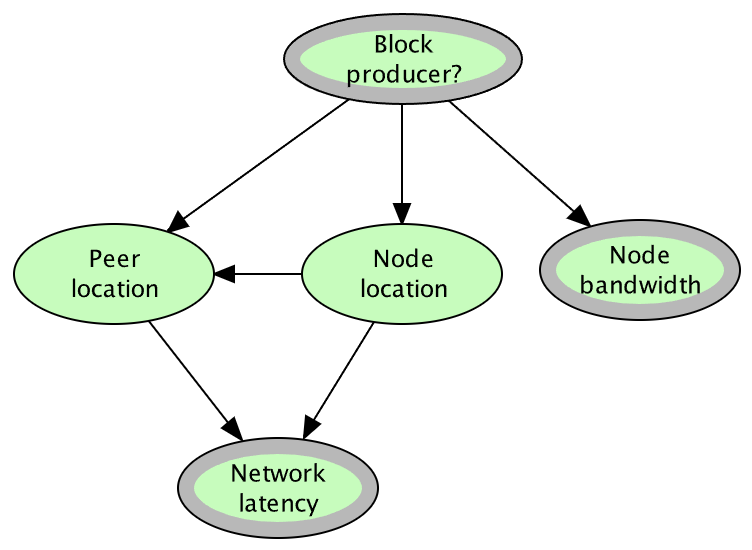}
\caption{Ethereum network BN }
\label{fig:node}
\end{minipage}
\hspace{.03\linewidth}
\begin{minipage}{.52\linewidth}
 \includegraphics[width=\linewidth]{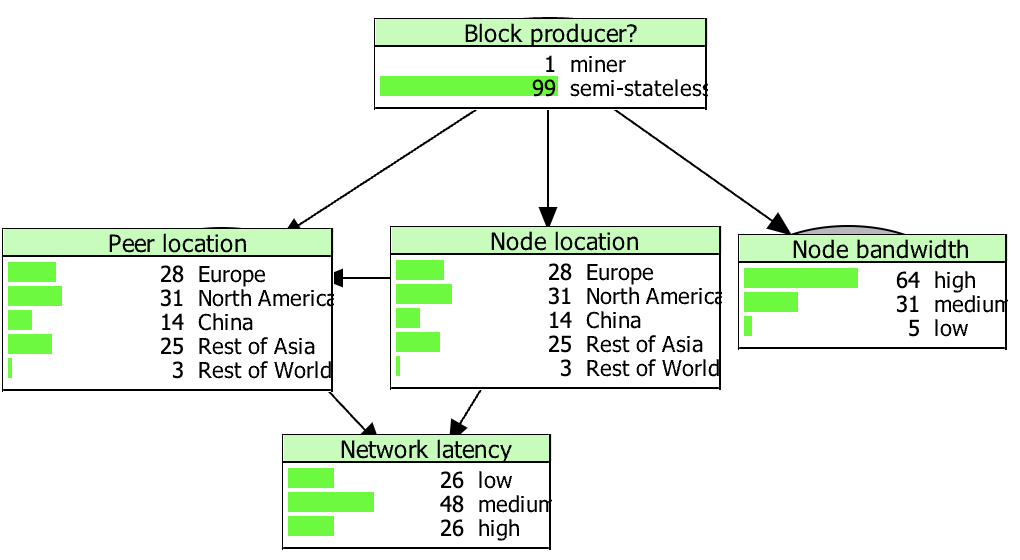}
\caption{ Running Ethereum network BN}
\label{fig:networkmonitors}
\end{minipage}
\end{figure}

The Ethereum network model (Figure \ref{fig:node} on page~\pageref{fig:node}) has five key factors: \textit{Block producer?}, \textit{Node bandwidth} and \textit{Network latency}, that are modelled as output `nodes' and \textit{Peer location} and \textit{Node location},  that are modelled as private nodes. 

For each of the factors in the model, we need a probability distribution across the various states that the factor can be in and this table of probabilities, known as a \gls{cpt}, is attached to each factor in the model.

 In our study we considered only two types of Ethereum nodes: those that produced validated blocks, i.e. `block producers', aka miners, and those that did not. We refer to the latter as `semi-stateless' nodes. For example, in Stateless Ethereum, a full node is treated as a `semi-stateless' node having varying degrees of `statelessness' from no state at the time of joining the network, up to `full state' if the node decides to keep up to date with the Ethereum world state. To quantify \textit{Block producer?} we calculated the proportion of all Ethereum nodes that are miners using data from Ethereum mainnet. This gave us an upper limit that was then adjusted using combined expert judgement. \textit{Node bandwidth} affects block propagation through the Ethereum network. Data from ethernodes.org \footnote{\url{https://ethernodes.org/} accessed 30 June 2020} were used to quantify this variable and a group of experts assisted in adjusting these proportions. 
 
Experts agreed on the threshold for ‘low’, ‘medium’, and ‘high’ and the description and ranges for these levels were documented with the elicitation sheets to ensure consistency and transparency. For example, for ‘node bandwidth’, businesses and hosting nodes were assigned as high bandwidth, colleges and residential nodes as medium bandwidth, and all other categories as low bandwidth.

Ethereum nodes are spread around the world and to make \textit{node location} tractable, we grouped countries into five regions: \textit{Europe}, \textit{North America}, \textit{China}, \textit{Rest of Asia}, and \textit{Rest of the world}. 

Five experts participated in the elicitation exercise for node location, using a range of strategies to estimate the probability distributions for mining and semi-stateless nodes, including ranking the locations and using the data from ethernodes.org as a baseline to vary for miners. {Peer location} was more challenging to quantify and we had to rely entirely on experts with domain knowledge to quantify this variable.

To quantify Network latency we used GNU ping data from Wondernetwork  \cite{WonderNetwork.2020}  grouped into the five regions used for node and peer locations. Running the Ethereum network \gls{oobn} model shows the marginal probabilities for each factor (Figure \ref{fig:networkmonitors} page~\pageref{fig:networkmonitors}).

\paragraph{Block creation \gls{oobn}}
\label{sec:blockcreationoobn}

Block creation in Ethereum is a complicated process, however, we found that many of the complexities of block creation did not directly affect the ability of stateless nodes to keep up with the network's activities. We built the initial version of the block creation model structure based on our understanding of the current block creation process in Ethereum, extracting data from 26,595 blocks that were mined during 26-30 November 2019. 
In Ethereum, the term gas is used as measure of the expected work required to process a transaction. Transactions are included in a block, but the total gas that they are expected to use is limited by the \textit{Block gas limit}, which can be adjusted by miners.

\begin{figure}
\centering
\begin{minipage}{.35\linewidth}
  \includegraphics[width=\linewidth]{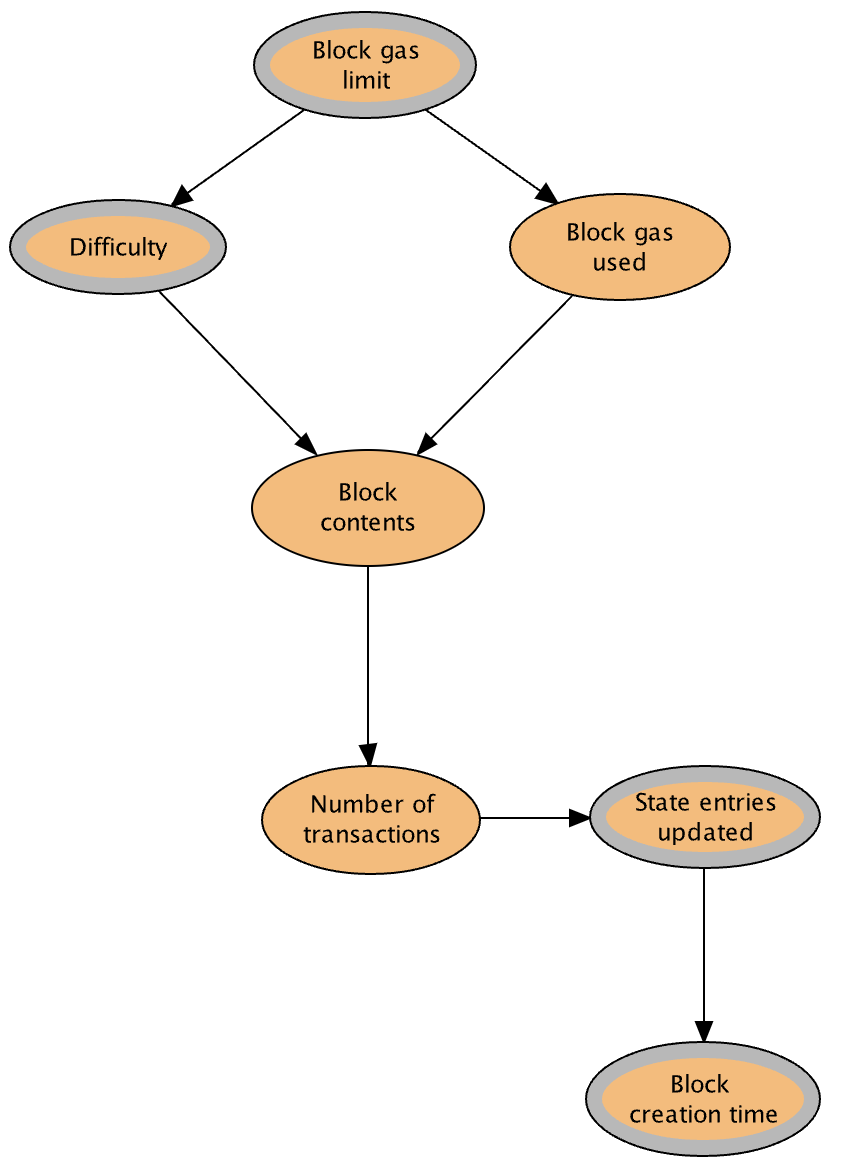}
\caption{Block creation BN }
\label{fig:blockcreation}
\end{minipage}
\hspace{.03\linewidth}
\begin{minipage}{.45\linewidth}
  \includegraphics[width=\linewidth]{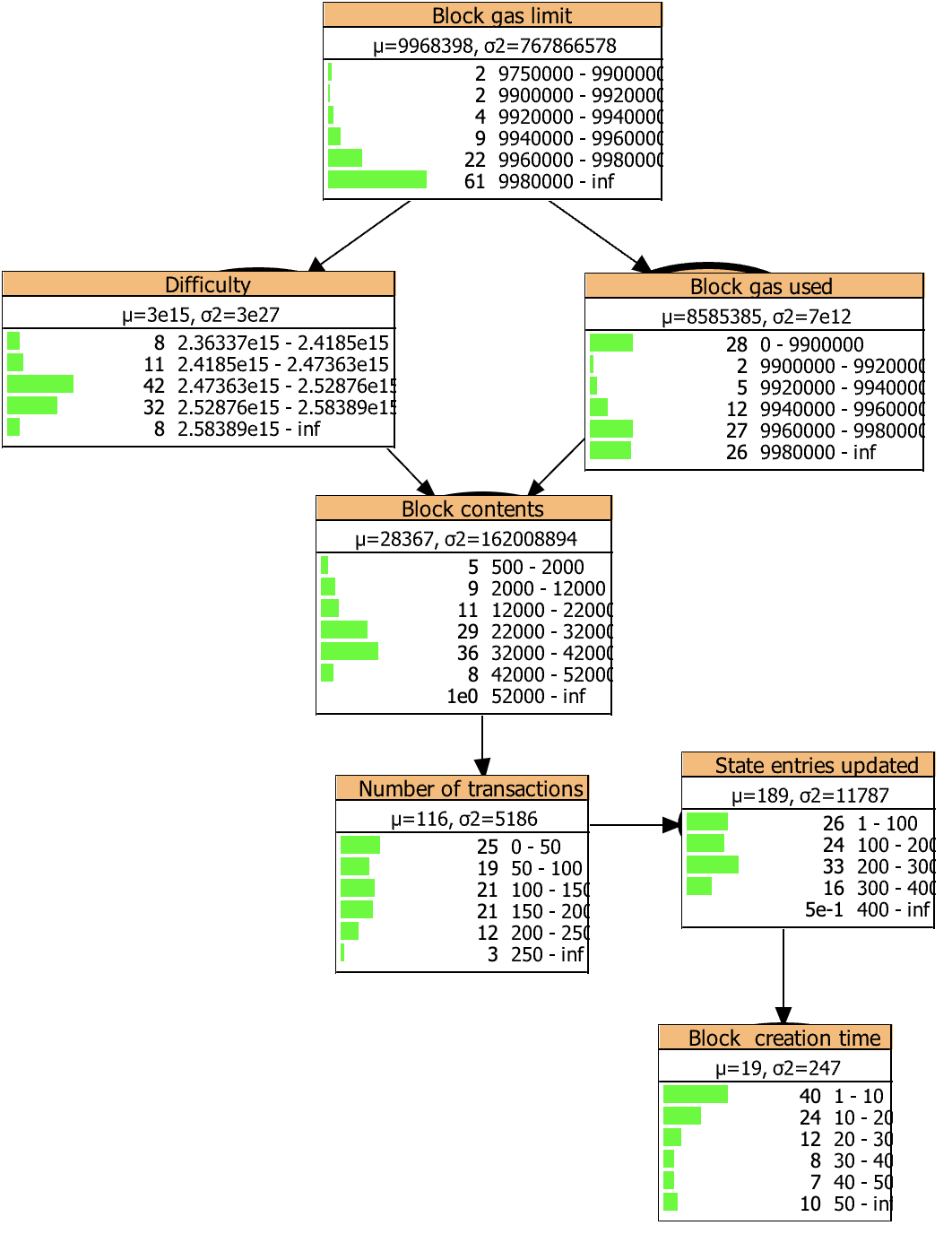}
\caption{ Running Block creation BN}
\label{fig:blockcreationmonitors}
\end{minipage}
\end{figure}

As can be seen in Figure  \ref{fig:blockcreation}, we read data such as block creation time, number of transactions per block, difficulty and more. We chose these blocks because we created witnesses for them when we implemented the Stateless witness specification.

Using the block information we extracted, we ran multiple structure learning algorithms using general and structure restricted methods. These algorithms are pre-coded in Hugin Expert\footnote{\url{https://www.hugin.com/}}, the software modelling tool we used to develop the model. The output from the learning exercises resulted in minor modifications to the initial structure. For example, although there was no direct link between \textit{Difficulty} and \textit{Block gas limit} in the way difficulty was calculated, three of the six structure learning algorithms picked up an association between them. The learning algorithms we used were: greedy-search-and-score, tree-augmented na\"{i}ve bayes, NPC (necessary path condition), Rebane-Pearl, a variant of the PC algorithm, and Chow-Liu. \footnote{\href{https://download.hugin.com/webdocs/manuals/GUI/pages/Manual/index.html\#algorithms}{Algorithms available in Hugin}}. We therefore supplemented the model with a directed link between these two key factors. However, weak links between key factors may be deleted in subsequent iterations of the model.
Figure \ref{fig:blockcreation} on page \pageref{fig:blockcreation} shows the resulting block creation model structure.

Running the block creation model, produces the marginal probability distributions as illustrated in Figure \ref{fig:blockcreationmonitors} on page \pageref{fig:blockcreationmonitors}.

 \paragraph{Witness creation \gls{oobn}}

\begin{figure}
\centering
\begin{minipage}{.32\linewidth}
 \includegraphics[width=\linewidth]{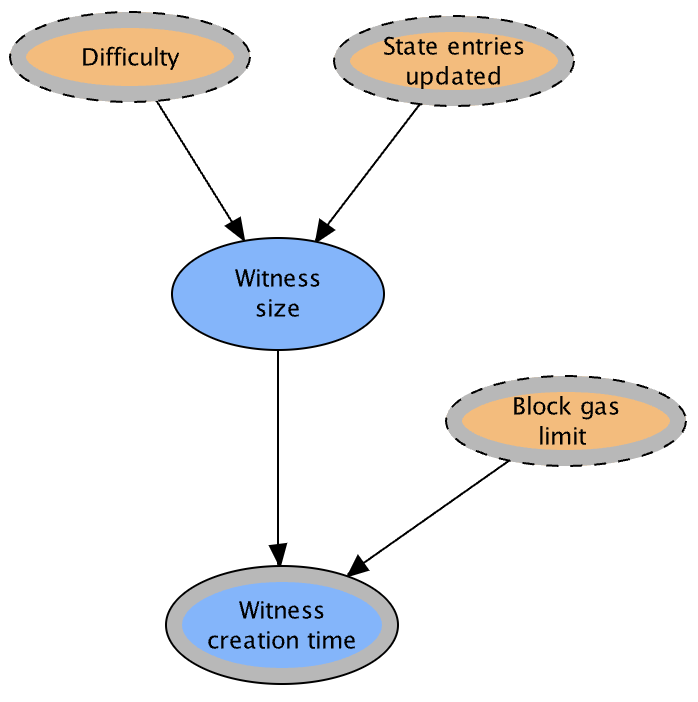}
\caption{Witness creation BN }
\label{fig:witness}
\end{minipage}
\hspace{.03\linewidth}
\begin{minipage}{.48\linewidth}
 \includegraphics[width=\linewidth]{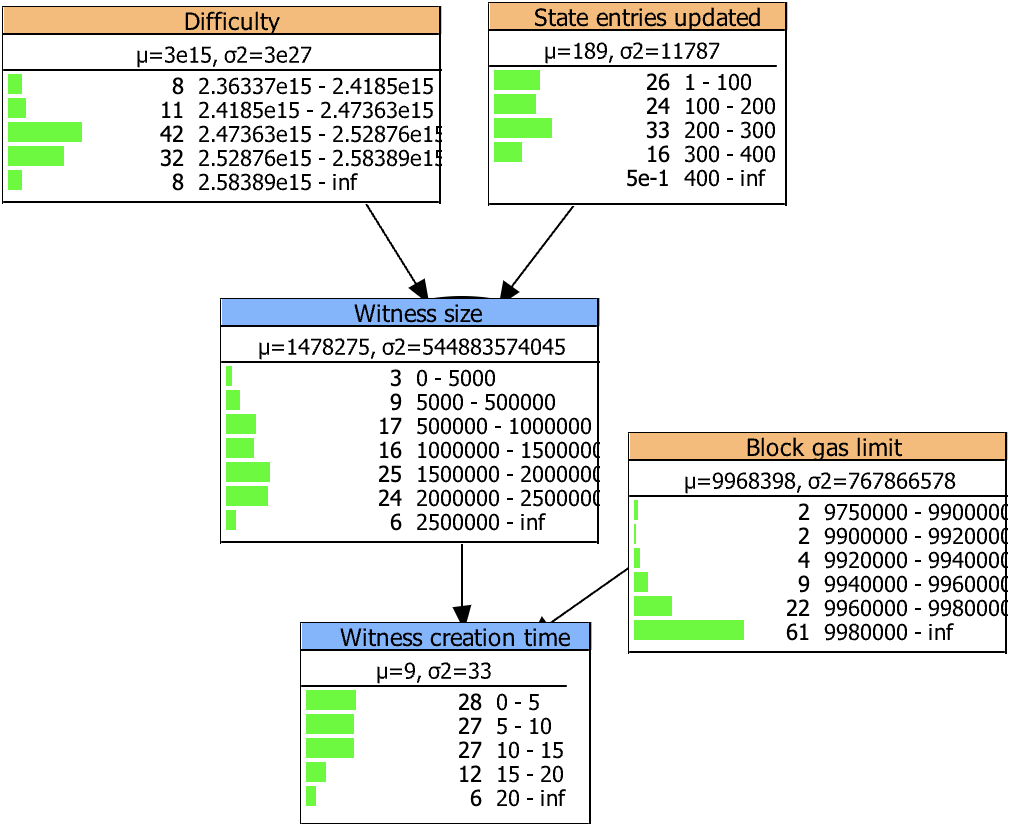}
\caption{Running Witness creation BN}
\label{fig:witnessmonitors}
\end{minipage}
\end{figure}

We implemented the Stateless Ethereum witness specification in a fork of Hyperledger Besu\footnote{\href{https://github.com/wcgcyx/besu/tree/witness-jsonrpc}{Implementation of Stateless Ethereum witness specification in fork of Hyperledger Besu}}, creating witnesses for 26,595 blocks, recording the size and creation time for each witness. The dependencies between witnesses and blocks were established using a combination of expert judgement and structure learning (Figure \ref{fig:witness} on page \pageref{fig:witness}).

Once the structure was ratified, we used the witness and block information to learn the conditional probabilities.

\textit{Difficulty} and \textit{State entries updated} variables are input nodes, obtained from the Block creation \gls{oobn}. \textit{Witness size} is private, and \textit{Witness creation time} is an output node, and thus visible to the other \gls{oobn} sub-models. 

\begin{figure}
\centering
\begin{minipage}{.46\linewidth}
 \includegraphics[width=\linewidth]{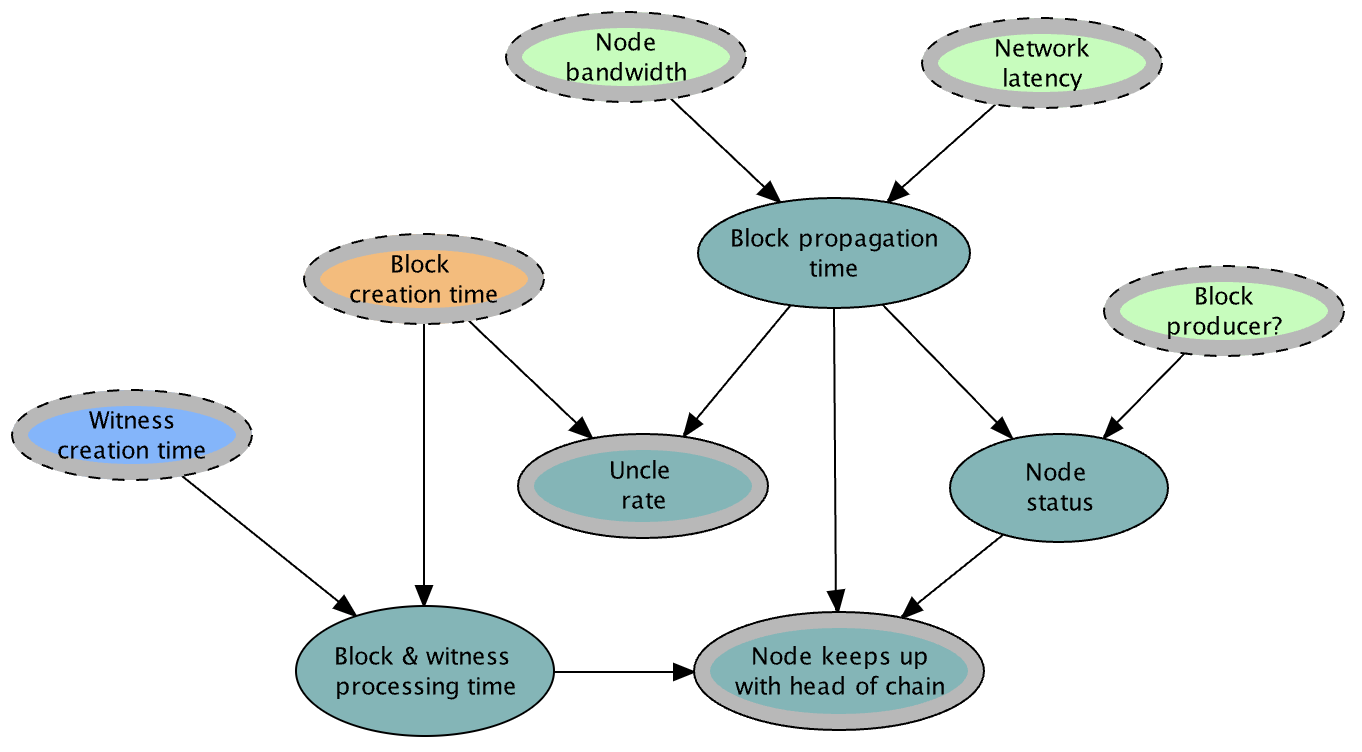}
\caption{Block propagation \gls{oobn} sub-model}
\label{fig:propagation}
\end{minipage}
\hspace{.01\linewidth}
\begin{minipage}{.51\linewidth}
 \includegraphics[width=\linewidth]{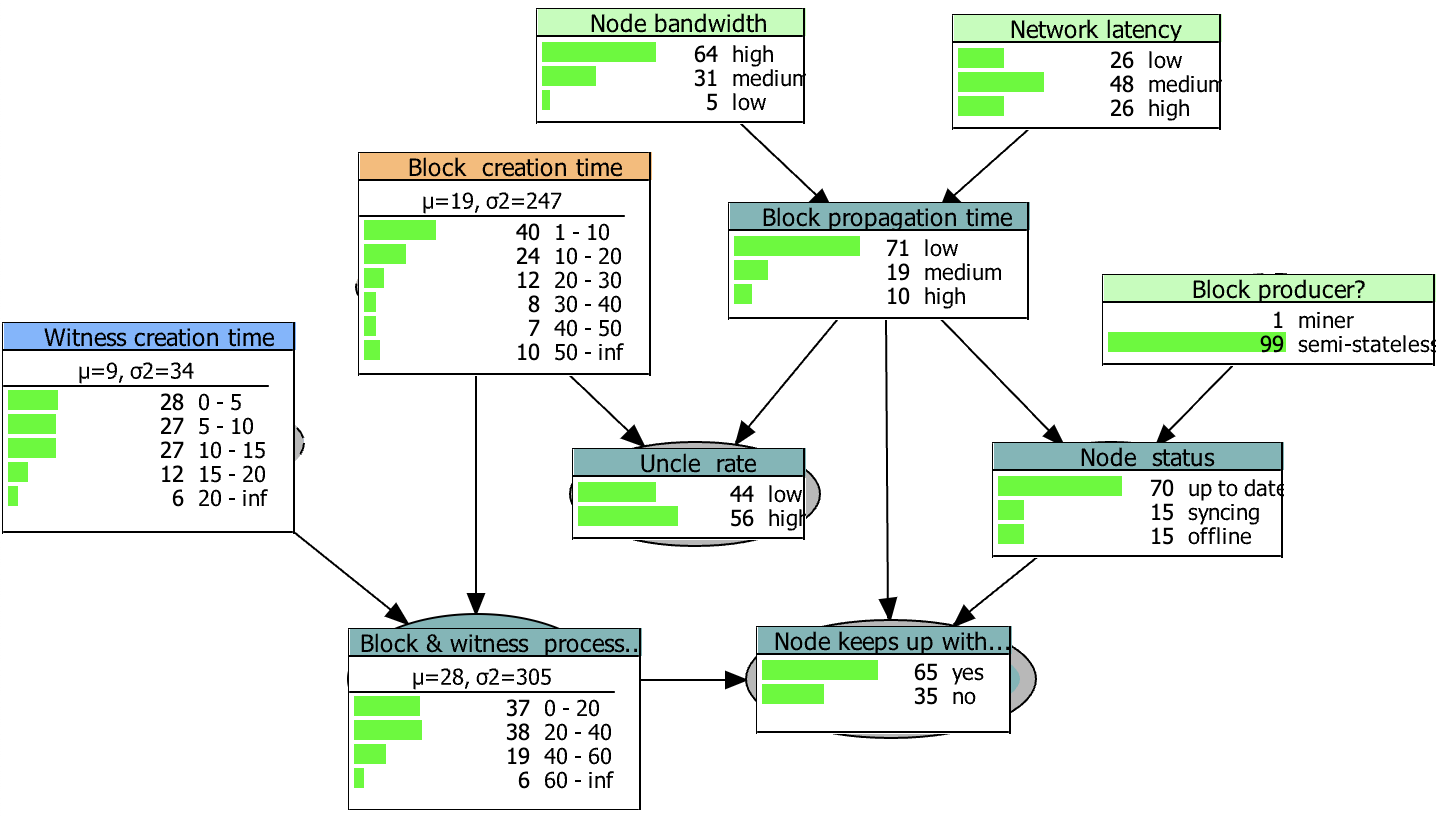}
\caption{Running Block propagation BN}
\label{fig:propagationmonitors}
\end{minipage}
\end{figure}

 \paragraph{Block Propagation \gls{oobn}}
\label{sec:statelessethereumblockpropagation}
The main model endpoint of Block Propagation is the variable \textit{Node keeps up with head of the chain}, which is modelled as an output node. Five of the factors in the block propagation BN model are part of other Stateless Ethereum sub-models. These are shown in Figure \ref{fig:propagation} on page \pageref{fig:propagation} as ellipses with a broken line: \textit{Block creation time}, \textit{Witness creation time}, \textit{Node bandwidth}, \textit{Network latency} and \textit{Block producer?}. We therefore only need to create CPTs for the remaining five nodes: \textit{Uncle rate}, \textit{Block propagation time}, \textit{Block and witness processing time}, \textit{Node status} and \textit{Node keeps up with head of the chain}.

The \textit{Node status} can be `up to date' (i.e. tracking the head of the chain), busy syncing to the head of the chain, or offline. When a node is offline it is difficult to know whether it is having difficulty keeping up to date with the head of the chain, has been decommissioned, having technical issues, or due to some other reason. We consulted with experts to obtain the conditional probabilities for \textit{Node keeps up with head of the chain} and the resulting CPT entries reflect their combined input.

When two or more miners produce a valid block at the same time and attempt to add it to the chain, only one will be added to the canonical chain. The other valid blocks are known as `uncles', or `ommers'. We calculated daily uncle rates recorded by etherchain.org\footnote{\url{https://etherchain.org/correlations}} and Alethio\footnote{\url{https://reports.aleth.io/}}. 

We collected block propagation times from ethstats.io\footnote{\url{https://ethstats.io/}} and ethstats.net\footnote{\url{https://ethstats.net/}}. Contribution of block propagation times to these websites are optional and do not represent all of the Ethereum network. Silva et al (2020) \cite{silva2020impact} conducted a month-long experiment from 1 April 2019 to 2 May 2019, collecting Ethereum network data, including block propagation times between geographically diverse nodes. We combined information from these three data sources for the prior probabilities, taking into account that changes in node latency and bandwidth will have an effect on block propagation time, which in turn has an effect on uncle rates. The extent to which the various factors influence block propagation time were elicited from experts.

\textit{Block and witness processing time} is calculated as the sum of \textit{Witness creation time} and \textit{Block creation time}. The assumption therefore is that these times are additive.

To quantify the effect that block propagation times have on \textit{Node status} for mining and non-mining nodes in the network we have to estimate the probability that a node is offline, syncing, or updated. Experts used our primary empirical data source, ethernodes.org, as a baseline to estimate the \gls{cpt} entries.

Running the fully quantified block propagation model results in the probabilities shown in Figure \ref{fig:propagationmonitors}. 

\begin{figure}
\centering
\begin{minipage}{.47\linewidth}
 \includegraphics[width=\linewidth]{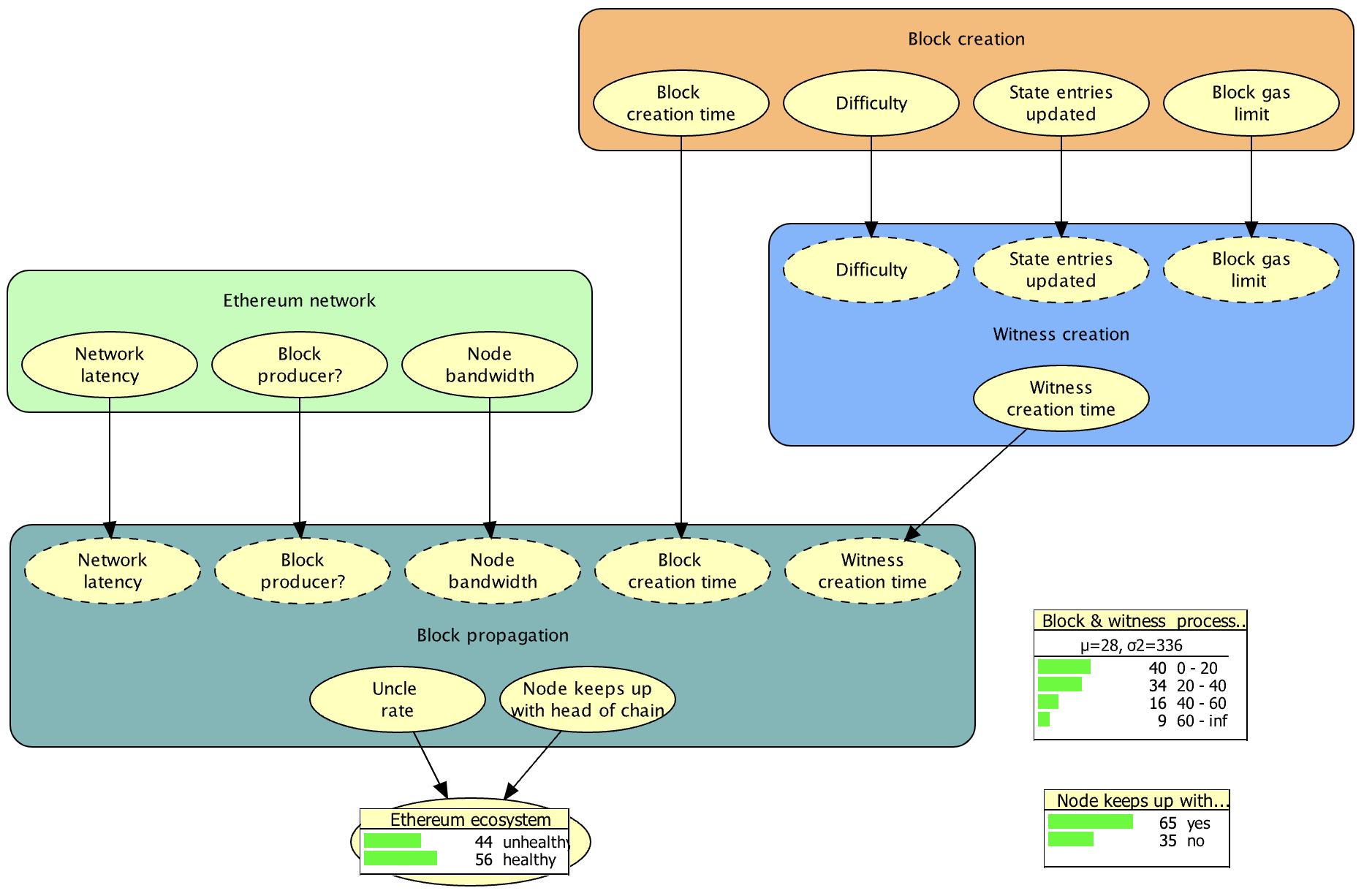}
\caption{Running Stateless Ethereum BN}
\label{fig:statelessmonitors}
\end{minipage}
\hspace{.01\linewidth}
\begin{minipage}{.48\linewidth}
 \includegraphics[width=\linewidth]{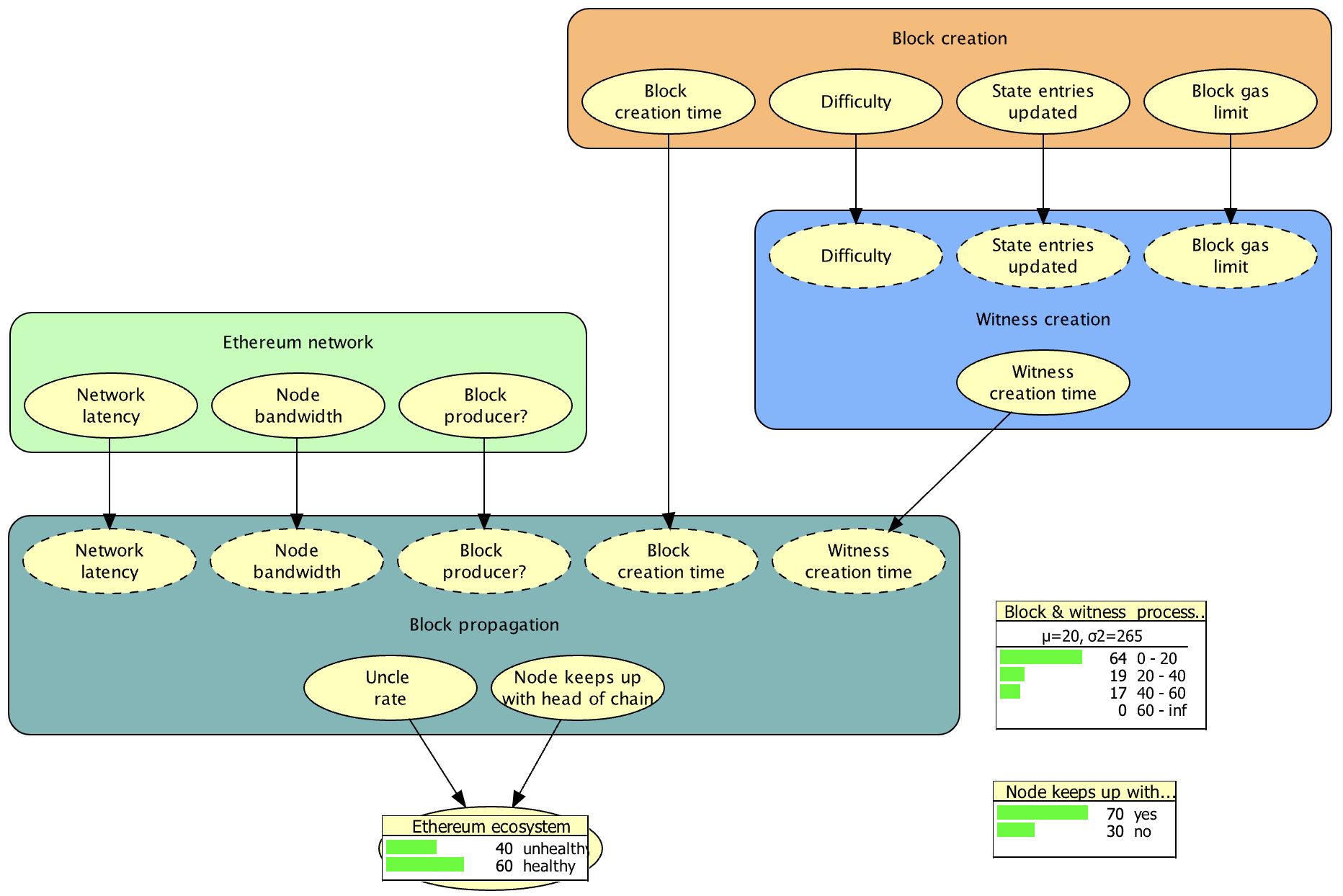}
\caption{Running Stateless Ethereum BN - no witness generation}
\label{fig:statelessnowitness}
\end{minipage}
\end{figure}

\paragraph{Ethereum ecosystem}
Whether \textit{Uncle rate} and \textit{Node keeps up with head of the chain} are good indicators of the health of the Ethereum ecosystem is open to debate, but the consensus among experts was that they are valuable indicators. The reasoning is that a high uncle rate is undesirable and indicative of a suboptimal network, adversely affecting transactional throughput\footnote{\href{https://medium.com/whiteblock/how-do-uncle-blocks-affect-blockchain-performance-9ce43c958772}{How do uncle blocks affect blockchain performance}}. Moreover, many nodes having difficulty synchronising to the network is indicative of network-related issues.
(Figure \ref{fig:statelessmonitors}). 

With the quantification of all the sub-models and the main model endpoint, \textit{Ethereum ecosystem}, complete, we were able to run the combined model, which ran all the sub-models (Figure \ref{fig:statelessmonitors}, page \pageref{fig:statelessmonitors}). It showed that the higher probability was the `healthy' state at 56\%. In other words, the health of the Ethereum ecosystem with a basic Stateless implementation, i.e. no compression of witnesses, is expected to be healthy. 

As a comparison, we ran the model again, excluding witness generation (Figure \ref{fig:statelessnowitness}, page \pageref{fig:statelessnowitness}). For this scenario there was a relative percentage gain of 7\% for the Ethereum ecosystem being healthy (60\%). 

The caveat for this result is three-fold: Firstly, the model was quantified using a subset of Ethereum mainnet data (26,595 blocks), which predates Ethereum improvement proposal EIP1559\footnote{\url{https://github.com/ethereum/EIPs/blob/master/EIPS/eip-1559.md}}. EIP1559 redesigned Ethereum's fee market. Of particular relevance to the Stateless Ethereum BN model is that post-EIP1559 block sizes are more variable and can be up to twice as large as they were at the time of building the Stateless Ethereum model. Secondly, implementing a witness size reduction technique such as Verkle tries\footnote{\url{https://dankradfeist.de/ethereum/2021/06/18/verkle-trie-for-eth1.html}} is bound to alter the model. Lastly, expert knowledge can introduce bias into the model.

For the reasons mentioned above, the probability of the final outcome, Ethereum ecosystem, being healthy, should not be interpreted as an exact result, but rather as a reference point for assessing changes in probability when exploring the effect that various scenarios have on the predicted health of the ecosystem.

We can use the quantified model to ask questions such as: ``For a non-mining node and a very large witness, how does that affect the ability of a node to keep up with the head of the chain?'' This will give some insight into expected changes to the Ethereum system status quo. 

\begin{figure}
\centering
\begin{minipage}{.25\linewidth}
  \includegraphics[width=\linewidth]{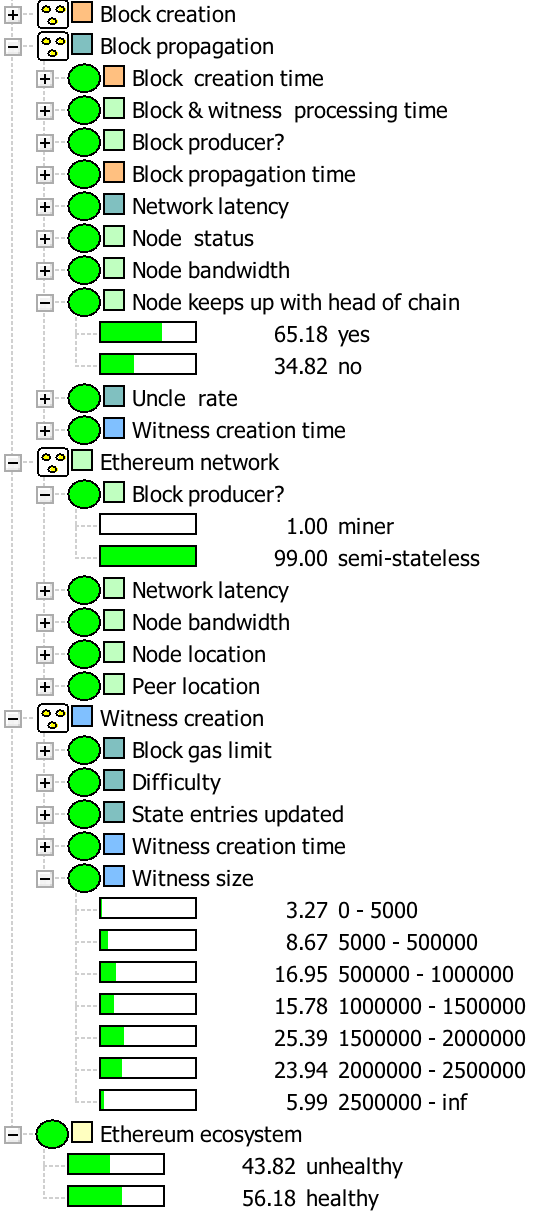}
\textbf{(a)}
\end{minipage}
\hspace{.01\linewidth}
\begin{minipage}{.25\linewidth}
  \includegraphics[width=\linewidth]{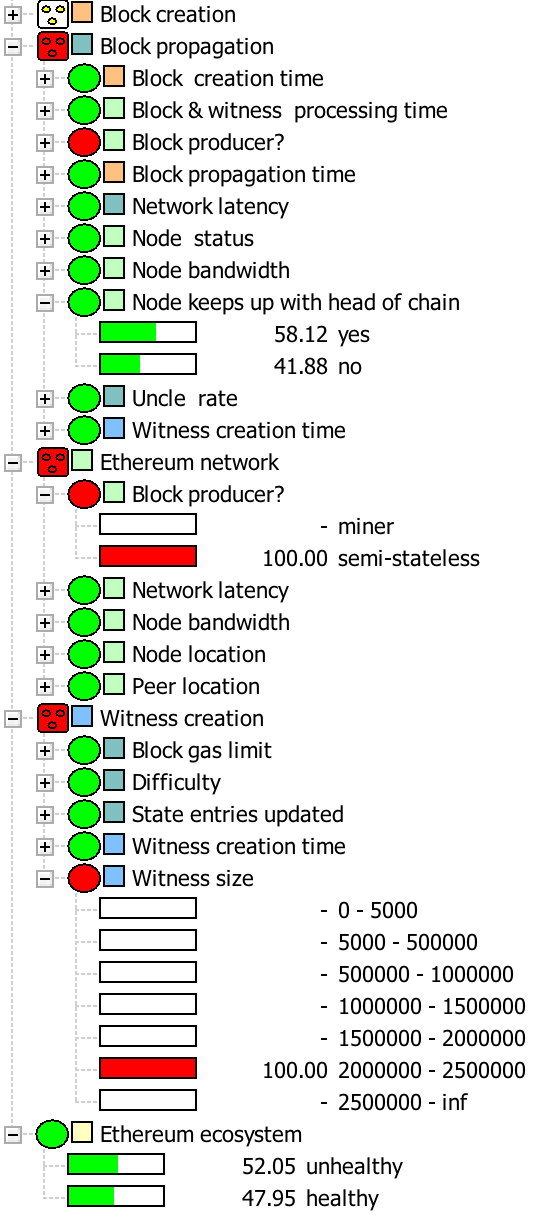}
\textbf{(b)}
\end{minipage}
\hspace{.01\linewidth}
\begin{minipage}{.25\linewidth}
  \includegraphics[width=\linewidth]{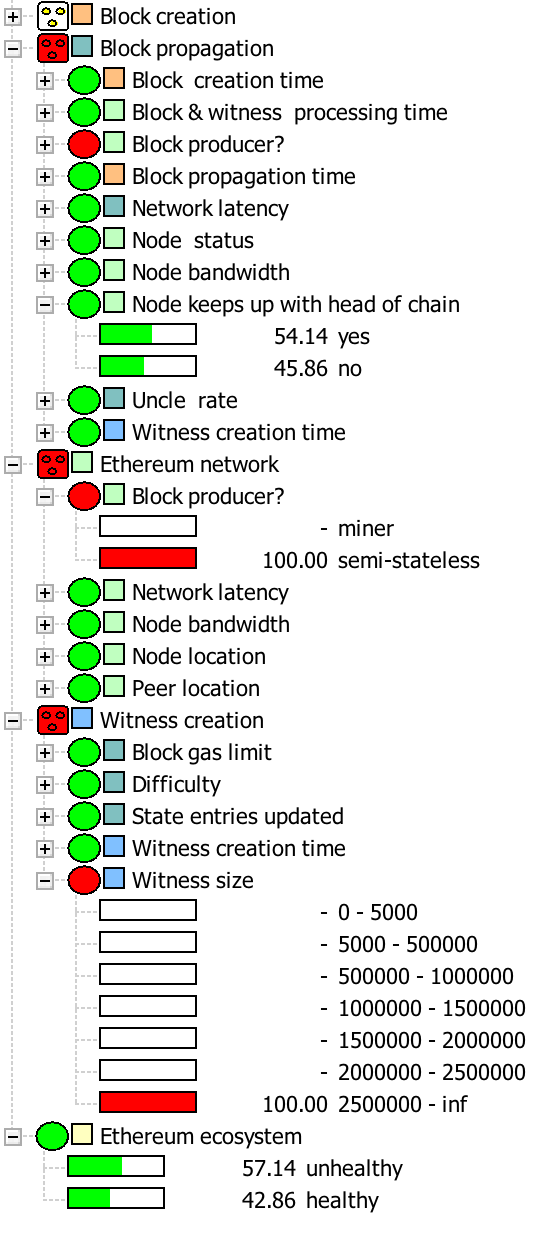}
\textbf{(c)}
\end{minipage}
\caption{Running `what if' scenarios in Stateless Ethereum OOBN}
\label{fig:scenarios}
\end{figure}

We selected two variations on this question and the graphical outcomes are shown in Figure \ref{fig:scenarios}. We observed that the probability of having this combination of evidence is 23.7\% for the second largest range (Figure \ref{fig:scenarios}(b)), and 5.9\% for the largest (Figure \ref{fig:scenarios}(c)). In other words, these scenarios are not expected to happen very often.

The less severe scenario (Figure \ref{fig:scenarios}(b)) showed the probability of keeping up with the head of the chain dropping from 65\% to 58\%, a percentage change of 7\% which is a relative change of 11\%. For the more severe scenario (Figure \ref{fig:scenarios}(c)), i.e. the largest witnesses, the probability of keeping up with the head of the chain dropped even further to 54\%, which translates to a relative change of 17\%.

\section {Evidence and Parameter Sensitivity Analysis}
Two additional important reasons for selecting BNs as the modelling technique are the explicit representation of assumptions in the graphical structure of the BN, which is helpful in the expert knowledge elicitation process (relevant for both structure and probability parameter elicitation), and the ability to perform different types of sensitivity analysis on the model to assess the robustness of the model and the results of the model. In this section, we will consider both types of sensitivity analysis. 

\begin{figure}
\centering
\begin{minipage}{.47\linewidth}
 \includegraphics[width=\linewidth]{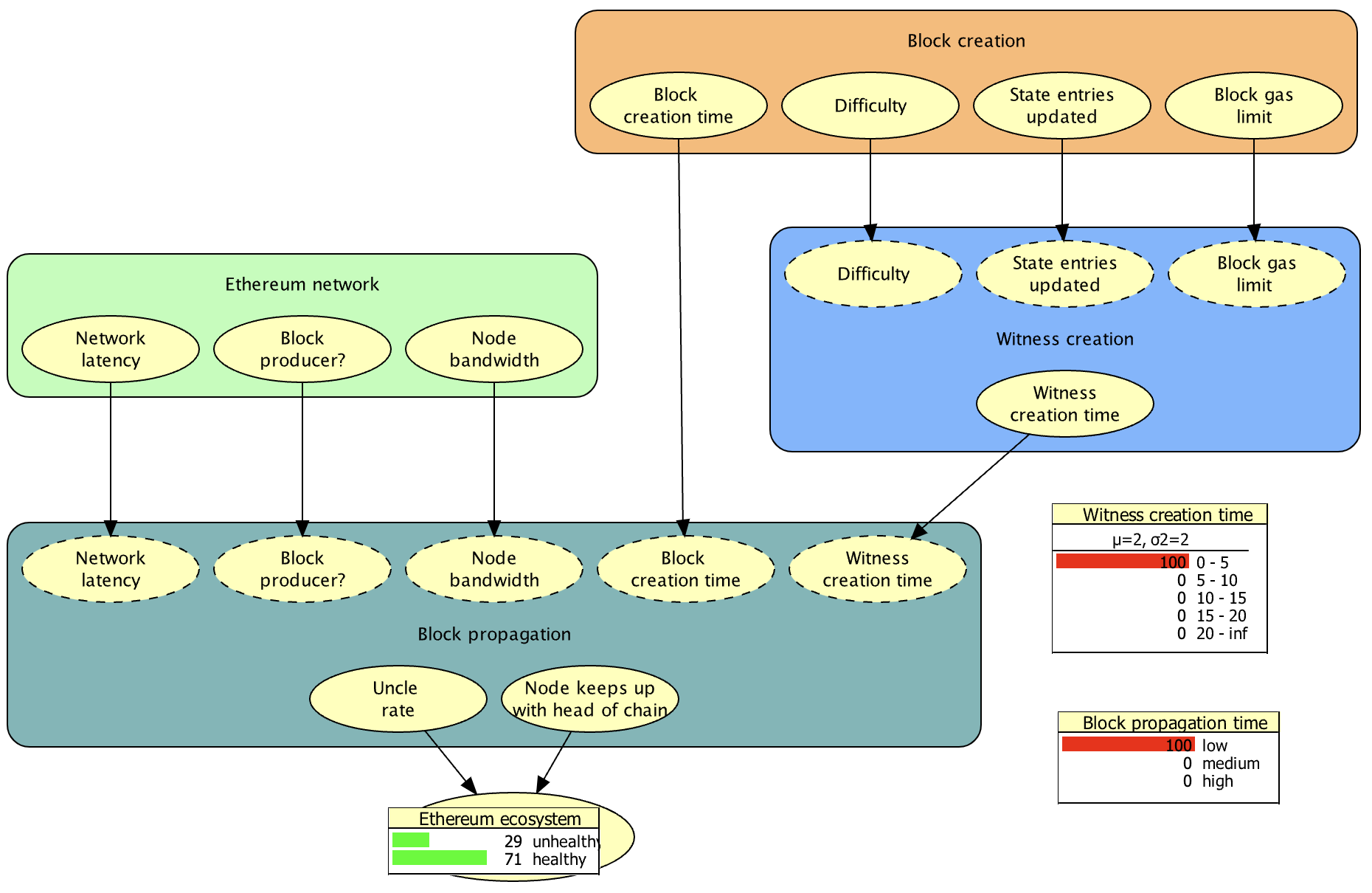}
\caption{Base scenario}
\label{fig:sensitivity1}
\end{minipage}
\hspace{.01\linewidth}
\begin{minipage}{.46\linewidth}
 \includegraphics[width=\linewidth]{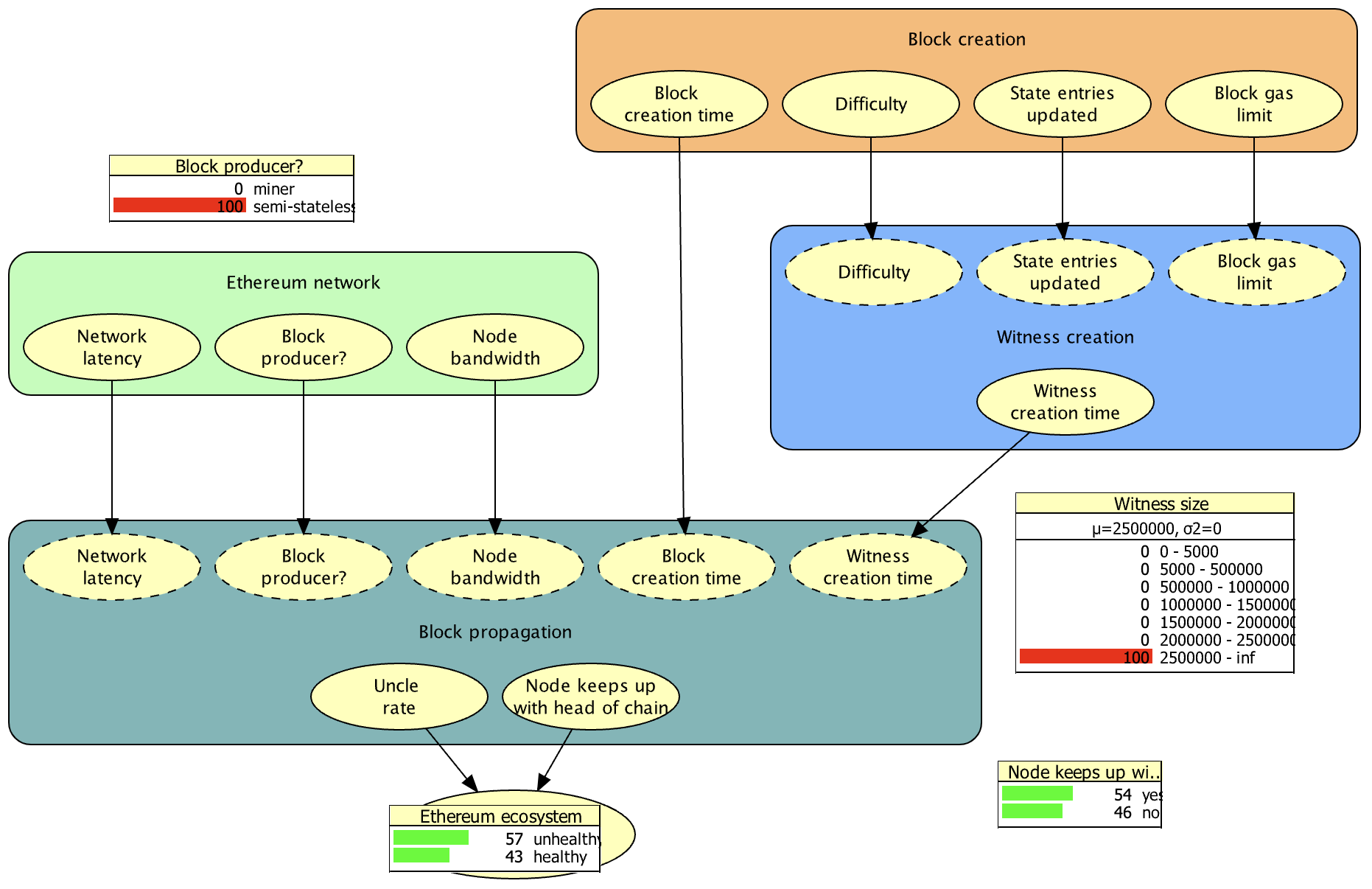}
\caption{Most severe scenario}
\label{fig:sensitivity2}
\end{minipage}
\end{figure}

The robustness of results of any model is sensitive to the underlying assumptions of the model. In a BN, one of the important assumptions is the quantification of the model, i.e., the entries of the CPTs also known as the parameters of the BN. To assess the robustness of the BN, a one-way parameter sensitivity analysis was performed under different evidence scenarios, see e.g.,~\cite{KjaerulffMadsen2013}. Here, we report on the parameter sensitivity analysis performed with respect to three different scenarios (1) no evidence, (2) base scenario, and (3) most severe scenario. The latter two scenarios are shown in Figure~\ref{fig:sensitivity1}  and Figure~\ref{fig:sensitivity2}, respectively. 

In each scenario, we investigated the impact of probability parameter variations on the probability of the hypothesis that the Ethereum ecosystem is healthy. We denote this as $P(EthereumEcosystem=healthy\mid \epsilon)$ where $\epsilon$ is the evidence scenario.
We performed a one-way parameter sensitivity evaluating the impact of variations in a single probability parameter $t$ on $P(EthereumEcosystem=healthy\mid \epsilon)$ determining the sensitivity function denoted $P(EthereumEcosystem=healthy\mid \epsilon)(t)$. 

In general, the sensitivity function is a ratio of two linear functions in the parameter, i.e., \\ 
$P(EthereumEcosystem=healthy\mid \epsilon)(t)=\frac{\alpha\cdot t + \beta }{\gamma \cdot t + \delta}$, where $\alpha, \beta, \gamma,$ and $\delta$ are real numbers. The composition of the evidence makes the denominator equal to one.

For the base scenario, the parameter with highest impact was \\
$t=P(EthereumEcosystem=healthy\mid \textit{NodeKeepsUpWithHeadOfChain}=yes,UncleRate=high)$, \\
i.e., this is the parameter with the highest value of the sensitivity function at the initial parameter assessment. 

The sensitivity function evaluated to $P(EthereumEcosystem=healthy\mid \epsilon)(t) = 0.3285\cdot t + 0.3318$ with a sensitivity value less than one. Hence, the probability of the hypothesis is relatively insensitive to variations in the parameter value.  \\
\noindent
The sensitivity function for the base case was: \\
$P(EthereumEcosystem=healthy\mid \epsilon)(t) = -0.7301\cdot t + 0.8222$, where \\$\textit{t=P(NodeStatus = stateOffline} \mid \textit{BlockPropagationTime = low,  EthereumNodeType = semiStateless)}$\\
\noindent
The sensitivity function for the most severe case was: \\
$P(EthereumEcosystem=healthy\mid \epsilon)(t) = -0.4724\cdot t + 0.6123$, where \\$\textit{t=P(NodeStatus = stateOffline} \mid BlockPropagationTime = low,  EthereumNodeType = semiStateless)$.

Again, the sensitivity value was less than one and in all three scenarios the sensitivity function was linear due to the structure of the evidence. This suggests that the posterior probability of the hypothesis is relatively insensitive to variations  in the probability parameter assessment when considered in a one-way parameter sensitivity analysis.

To further assess the performance of the model an evidence sensitivity analysis was performed. In evidence sensitivity analysis, the robustness of the results of the model was assessed with respect to variations in the evidence, see, e.g.,~\cite{KjaerulffMadsen2013}.  For the no evidence scenario, the two variables that may produce the highest impact on the probability of the hypothesis, are \textit{NodeKeepsUpWithHeadOfChain} and  $NodeStatus$ that may produce $0.0337 \leq P(EthereumEcosystem=healthy\mid \epsilon)\leq 0.8439$ and  $0.0536 \leq P(EthereumEcosystem=healthy\mid \epsilon)\leq 0.6764$, respectively.  

For the base scenario, the same two variables  may produce $0.06 \leq P(EthereumEcosystem=healthy\mid \epsilon)\leq 0.91$ and  $0.08 \leq P(EthereumEcosystem=healthy\mid \epsilon)\leq 0.83$, respectively.  For the most severe scenario,  two variables  may produce $0.0176 \leq P(EthereumEcosystem=healthy\mid \epsilon)\leq 0.7768$ and  $0.0286 \leq P(EthereumEcosystem=healthy\mid \epsilon)\leq 0.5271$, respectively.  The results of the evidence sensitivity analysis showed that the two variables that have the highest impact on the hypothesis across the three different evidence scenarios were \textit{NodeKeepsUpWithHeadOfChain} and  \textit{NodeStatus}. All other variables can produce a much smaller variation in the probability of the hypothesis.

\section{Discussion}
Introducing Stateless Ethereum into a functioning, stable ecosystem presents us with an interesting study. 

Ethereum has an active open source community which, by its very nature, encourages collaboration among developers and researchers. Moreover, Ethereum mainnet is a public blockchain, giving us ready access to empirical data for the BN model we built. However, contribution to websites reporting on Ethereum nodes and the distributed network is voluntary, and therefore, any insights from the data are based on a partial view of the overall network. 

Modelling the known processes and quantifying them using empirical data, supplemented with the new processes and representing our knowledge about these processes, enables us to gain a more complete picture of the potential repercussions to other parts of the network. The ability of Bayesian networks to include diverse data sources such as empirical data, model output and expert knowledge, provides us with a view of the proposed system that takes into account all available information at the present time. However, the presence of expert knowledge can introduce bias into the model. Conducting sensitivity analysis and model verification are helpful in identifying and mitigating bias. Although not always the case, in this instance, the sensitivity analysis performed to assess the robustness of the model did not reveal any unexpected model behaviour requiring a revision of the model structure. However, since variables \textit{NodeKeepsUpWithHeadOfChain} and \textit{NodeStatus} have the highest impact on the hypothesis that the Ethereum ecosystem is healthy, additional effort and validation in estimating those parameters are recommended \cite{Johnson2013}.

Bayesian network modelling is typically an iterative process, and the model presented here is the first iteration of the model. Additional versions may be created to include new and emerging research and current data, such as post-EIP1559.
A key challenge in the development of this probabilistic model and its quantification has been the varying degree of data coverage of the datasets used to populate the \glspl{cpt} of the \gls{bn} nodes. This inconsistency introduces measurement errors into the model, compounded by elicited probabilities and interpolation, which introduce bias into the model. We advise that the reader take this into consideration when assessing the predicted probabilities from running the various scenarios. \glspl{bn} are particularly useful in providing a statistically sound approach to representing the initial state of a system and identifying those areas that would benefit most from additional data collection and further in-depth statistical and mathematical modelling. 

Although the main objective and aim of Stateless Ethereum remain firm, the ways in which this can be achieved has evolved over time, with some changes standing the test of time, and others being assessed and discarded in favour of more pragmatic proposals. Even at this late stage of development, a major new initiative has been introduced - \textit{Regenesis}, which is rapidly gaining traction in the Stateless Ethereum research community. These changing conditions make modelling more challenging, but these changes can be largely mitigated or minimised by the modular approach taken in building this model, as a new sub-model can be added in the next iteration of the model to represent the changes introduced by regenesis. Moreover, as we focus our attention of data collection to those areas that have a greater impact on the probability of a node keeping up with the network, we will refine this model over time, learning from new data and information as they become available. 

An online, interactive version of this BN model is hosted by Hugin on their Demo website\footnote{\url{https://demo.hugin.com/example/StatelessEthereumModel}}. Furthermore, three blog posts describe the model, providing additional detail, not included in this paper.\footnote{\href{https://consensys.net/blog/research-development/modelling-stateless-ethereum-a-journey-into-the-unknown/}{Defining Stateless Ethereum: A Journey Into The Unknown}} \footnote{\href{https://consensys.net/blog/research-development/building-a-stateless-ethereum-model/}{Building A Stateless Ethereum Model}} \footnote{\href{https://consensys.net/blog/research-development/measuring-the-health-of-the-ecosystem-in-a-stateless-ethereum/}{Measuring The Health Of A Stateless Ethereum Ecosystem}} 

\bibliographystyle{eptcs}
\bibliography{stateless}
\end{document}